\documentclass[aps, pra, twocolumn, floatfix]{revtex4-1}
\pdfoutput=1
\usepackage{amsmath}
\usepackage{bm}
\usepackage{graphicx}
\usepackage[pdfusetitle]{hyperref}
\usepackage[capitalise]{cleveref}

\newcommand{\Tmat}{\mathcal{T}}
\newcommand{\F}{\textnormal{F}}
\newcommand{\redmass}{m_{r}}
\newcommand{\dif}{\textnormal{d}}
\newcommand{\voldif}[1]{\dif^{3} #1}

\begin{document}

\title{Quasiparticle scattering rate in a strongly  polarised  Fermi mixture}

\author{Rasmus S\o gaard Christensen}
\affiliation{Department of Physics and Astronomy, University of Aarhus, Ny Munkegade, DK-8000 Aarhus C, Denmark}
\author{Georg M. Bruun}
\affiliation{Department of Physics and Astronomy, University of Aarhus, Ny Munkegade, DK-8000 Aarhus C, Denmark}

\begin{abstract}
We analyse the scattering rate of an impurity atom in a Fermi sea as a function of momentum and temperature   in the BCS-BEC crossover. 
The  cross section is calculated
using a microscopic multichannel theory for the Feshbach resonance  scattering, including finite range and 
 medium effects. We show that pair  correlations significantly increase the cross section for strong interactions close to the unitarity regime.  
 These pair correlations give rise to a  molecule  pole  of the cross section at negative  energy on the BEC side of the resonance, which smoothly evolves
  into  a resonance  at positive scattering energy with a non-zero imaginary part on the BCS  side. 
 The resonance is the analogue of superfluid pairing for the corresponding population balanced system. Using Fermi liquid theory, we then show that the  
  low temperature   scattering rate of the impurity atom is  significantly increased due to these pair correlations for low momenta.  
  We demonstrate that finite range and mass imbalance effects are significant for the experimentally relevant $^6$Li-$^{40}$K mixture, 
and we finally discuss how the scattering rate can be measured using radio-frequency spectroscopy and Bose-Fermi mixtures. 
\end{abstract}

\maketitle

\section{Introduction}
The scattering between quasiparticles  plays a fundamental role for  our understanding of many-body systems. 
Fermi liquid theory is based on a 
 small quasiparticle scattering rate for low energies and temperatures~\cite{BaymPethick1991book}, and quasiparticle scattering determines the transport properties of quantum systems which is of great  
importance for modern and future information technology. 
The scattering  rate between electrons in  two-dimensional  quantum well structures has been  
measured~\cite{Berk1995,Murphy1995,Slutzky1996}, but there are 
 few experiments providing a clean measurement of the quasiparticle scattering rate, and 
 the inevitable presence of disorder and other complications makes a  
quantitative comparison with theory  challenging for solid state systems~\cite{Qian2005}.
A mobile impurity particle immersed in a medium with a continuous set of degrees of freedom 
provides a  clear realisation of a quasiparticle. In  seminal works, Landau and Pekar demonstrated 
 that electrons in a dielectric medium become dressed by the collective excitations of the material forming a quasiparticle called a polaron~\cite{Landau1933,Pekar1946}.
 The study of polaron physics 
  has  subsequently grown into an independent research field~\cite{Mahan2000book}. Other examples of  impurity 
  quasiparticles  include helium-3 mixed with helium-4~\cite{BaymPethick1991book}, and $\Lambda$ particles 
  in nuclear matter \cite{Bishop1973}.

A major step forward in the study of impurity quasiparticles was recently made with the creation of 
highly population imbalanced two-component cold atomic gases. When 
the population of one of the components (majority atoms) is much larger 
than the population of the other component (minority atoms), the minority atoms 
play the role of the mobile impurities and the majority atoms play the role of the medium \cite{Schirotzek2009,Kohstall2012,Koschorreck2012}. These experiments have lead 
to a renewed focus on impurity physics. A surprising result which has emerged from these studies is 
that the impurity atom forms a well-defined quasiparticle, coined the Fermi polaron, whose zero-momentum properties can be described accurately in terms of a rather simple theory, 
even for  large coupling strengths~\cite{Chevy2006, Lobo2006,Punk2009,Prokofev2008,Mathy2011,Massignan_Zaccanti_Bruun}. 
While  the polaron for zero momentum has been intensely studied both experimentally  and theoretically, 
much less attention has been given to its properties for a non-zero momentum. In particular, the finite 
life-time of a polaron with non-zero momentum
 due to scattering with the majority particles is almost 
exclusively studied in the weak coupling regime. The momentum relaxation rate of the polaron was calculated to second order in the interaction~\cite{Bruun2008}, and 
the imaginary part of the self-energy of the polaron
due to  collisions was calculated perturbatively to second order in the impurity-majority atom scattering length $a$ for a broad 
resonance at zero temperature~\cite{Trefzger_Castin,Trefzger2014}, as well as for temperatures much less than the Fermi temperature~\cite{Lan2013}. The
 high momentum case was analysed using the operator product expansion~\cite{Nishida2012}. 

Here, we calculate the scattering rate of an impurity atom (polaron) with non-zero momentum on
 the majority fermions using a non-perturbative approach to include 
strong coupling effects. The  Feshbach scattering between the impurity atom and the Fermi sea is described with a microscopic multichannel theory   including 
finite range and medium effects. We analyse in detail the energy and momentum dependence of the cross section for 
the scattering, and show that 
 medium effects significantly increase the cross section for strong interactions close to the unitarity regime.  
 This is due to pair correlations giving rise to a  molecule pole in the cross section at negative   energy in the BEC regime ($a>0)$, which evolves smoothly into 
  a resonance with 
 non-zero imaginary part at positive energy  in the BCS regime ($a<0$). This resonance  
 is the remnant of superfluid pairing in the equivalent system with equal population of the two components.     
We then calculate the collision rate of the polaron with the majority atoms as a function of momentum
 and temperature using Fermi liquid theory. We 
show that it is significantly increased  due to  pair 
correlations for energies and temperatures  comparable to or smaller than the Fermi energy 
$\epsilon_\F $ of the majority particles. The effects of a non-zero effective range and a difference in the masses 
of the impurity and majority atoms 
are examined throughout, and we show that they are significant for the experimentally relevant $^6$Li-$^{40}$K mixture. Finally, we discuss how the scattering rate can be 
measured experimentally using radio-frequency (RF) spectroscopy or Bose-Fermi mixtures. Such a measurement would present an important  step forward 
due to the relatively  few unambiguous  
experimental  results concerning the scattering rate of quasiparticles in the literature.

\section{Theory}
Consider  a single spin $\downarrow$ particle  interacting with a Fermi sea of spin $\uparrow$ particles forming a Fermi polaron.
In this paper, we investigate the collision rate  between the polaron and the majority cloud using Landau Fermi liquid theory. 
From Fermi liquid theory, the collision rate of a quasiparticle with momentum 	${\mathbf p}_1$ with the surrounding medium is~\cite{BaymPethick1991book}
\begin{multline}
\frac 1 {\tau_{p_1}} =
	\int \frac{ \voldif{p}_{2} }{ (2\pi)^3 } \frac{ \voldif{p}_3 }{ (2\pi)^3 }
		W \delta(\epsilon_{p_1\downarrow}+\epsilon_{p_2\uparrow}-\epsilon_{p_3\downarrow} - \epsilon_{p_4\uparrow})
	\\
		\times [f_{p_2\uparrow} (1 - f_{p_3\downarrow}) (1 - f_{p_4\uparrow}) - f_{p_3\downarrow} f_{p_4\uparrow} (1 - f_{p_2\uparrow} ) ],
\label{eqn:Lifetime}
\end{multline}
where  $\epsilon_{p s} = p^{2}/2m_{s}$, $m_s$ is the mass of the spin $s$ particles,  and $\bm{p}_{4} = \bm{p}_{1} + \bm{p}_{2} - \bm{p}_{3}$.
The transition probability for scattering of $\uparrow$ and $\downarrow$ particles with momenta  ${\mathbf p}_1$ and ${\mathbf p}_2$ respectively to 
momenta ${\mathbf p}_3$ and ${\mathbf p}_4$ is $W(12;34) = 2 \pi^2 \sigma/m_r^2$,
where $\sigma$ is the cross section and $m_r=m_\uparrow m_\downarrow/(m_\uparrow+m_\downarrow)$ is the reduced mass. 
 We use units where $\hbar=k_B=1$.
  The distribution functions are $f_{p s} = [ e^{\beta ( \epsilon_{p s} - \mu_{s}) } + 1 ]^{-1}$
with $\mu_{s}$ the chemical potential of the spin $s$  particles, 
and we can  as usual for atomic gases  assume
 that there is only $s$-wave scattering between $\downarrow$  and $\uparrow$ atoms, whereas there is no interaction between equal spin atoms. 
The focus of this paper is on medium effects on the scattering rate, and in particular the effects of pairing correlations. For 
 this purpose it is sufficient to use $\epsilon_{p\downarrow}=p^2/2m_{\downarrow}$  for the polaron energy and in addition assume that it has 
unit quasiparticle residue. This limits our analysis to the regime $k_\F |a|\lesssim 1$ where the polaron is well defined~\cite{Massignan_Zaccanti_Bruun}. Here,  
  $k_\F=\sqrt{2m_\uparrow\epsilon_\F}$ is the Fermi momentum.  As we shall see, there are still strong pair correlations leading to 
significant effects on the cross section in this regime. Medium effects on the quasiparticle energy and residue have been investigated intensively, see e.g.\ 
Ref.~\cite{Chevy2006, Lobo2006,Punk2009,Prokofev2008,Massignan_Zaccanti_Bruun}. 

Using energy/momentum conservation and the fact that $f_\downarrow=0$ in the polaron limit of a single $\downarrow$ atom, 
 \cref{eqn:Lifetime} can be simplified to
 \begin{equation}
\frac 1 {\tau_{p_1}}=\int\frac{d^3p_2}{(2\pi)^3}\frac{m_rp_r}{8\pi^3}\int d\Omega
Wf_{2\uparrow}(1-f_{4\uparrow}).
\label{Lifetime2a}
\end{equation}
Here $\Omega$ is the solid angle for the direction of the out-going relative momentum $\bm{p}'_{r} = (m_{\uparrow} \bm{p}_{3} - m_{\downarrow} \bm{p}_{4}) / M$
 with respect to the in-going relative momentum  $\bm{p}_{r} = (m_{\uparrow} \bm{p}_{1} - m_{\downarrow} \bm{p}_{2}) / M$ of the 
 scattering process, where $M = m_{\downarrow} + m_{\uparrow}$.

In order to   describe experimentally relevant atomic systems interacting via a Feshbach resonance, 
we use a multichannel theory for the atom-atom scattering which includes finite range effects. 
Medium effects on the scattering matrix are described using the ladder approximation, which  
has turned out to yield surprisingly accurate results for the Fermi polaron~\cite{Massignan_Zaccanti_Bruun}. The scattering 
matrix in the open channel can be written as~\cite{BruunKolomeitsev}
\begin{equation}
	\Tmat( P, \omega ) =
			\frac{ \Tmat_{\mathrm{bg}} }
				{ 
					\left( 1 + \frac{ \Delta \mu \Delta B }{ \tilde{\omega} - \Delta \mu ( B - B_{0} ) } \right)^{-1}
					- \Tmat_{\mathrm{bg}} \Pi( P, \omega )
				}
	\label{eqn:Tmat_init}
\end{equation}
with $\bm{P} = \bm{p}_{1} + \bm{p}_{2}$ the center of mass momentum, $\tilde{\omega} = \omega - P^{2}/2M$ is the energy in the relative frame, 
and  $\Tmat_{\mathrm{bg}} = 2 \pi a_{\mathrm{bg}} / \redmass$, 
where $a_{\textnormal{bg}}$ is the background  scattering length.  The width and position of the Feshbach resonance
is $\Delta B$ and $B_{0}$ respectively, and
the difference in magnetic moments between the Feshbach molecule and the pair of open channel scattering atoms
is $\Delta \mu $. The  pair propagator is
\begin{equation}
	\Pi( P, \omega ) =
					\int \frac{\voldif{q}}{(2 \pi)^{3}}
					\left(
						\frac{ 1  - f_{k' \uparrow} }{ \omega + i0_{+} - \frac{P^{2}}{2M} - \frac{q^{2}}{2 \redmass} }
						+ \frac{2 \redmass}{q^{2}}
					\right),
	\label{eqn:Pair_Propagator}
\end{equation}
where $\bm{k} = \frac{m_{\downarrow}}{M} \bm{P} + \bm{q}$ and $\bm{k}' = \frac{m_{\uparrow}}{M} \bm{P} - \bm{q}$. 
The cross-section for scattering with center-of-mass momentum ${\mathbf P}$ and relative momentum ${\mathbf p}_r$ is then 
given by the on-shell scattering matrix as $\sigma=m_r^2|\Tmat(P,P^2/2M+p_r^2/2m_r)|^2/\pi$.
In a vacuum, 
we have $\Pi_{\mathrm{vac}}( P,\omega ) = - i \redmass^{3/2} \sqrt{ \omega-P^2/2M } / \sqrt{2} \pi$, and the on-shell 
 ${\mathcal T}$-matrix \cref{eqn:Tmat_init} becomes 
\begin{align}
	\Tmat_{\rm vac} =
			\frac{ 2 \pi a / \redmass }
				{
					\frac{ p_r^2 r_{\mathrm{eff}}  ( a - a_{\mathrm{bg}} )/2 - 1 }
						 { p_r^2 r_{\mathrm{eff}} a_{\mathrm{bg}} ( 1 - a_{\mathrm{bg}} / a )/2 - 1 }
					+ i p_{r} a
				}.
	\label{eqn:Tmat_vac}
\end{align}
The effective range is  $r_{\mathrm{eff}} = - ( \redmass a_{\mathrm{bg}} \Delta \mu \Delta B )^{-1}$ and 
we have used the usual relation $a = a_{\mathrm{bg}} [ 1 - \Delta B / (B - B_{0}) ]$.
In the limit $p_{r} r_{\mathrm{eff}} \ll 1$, \cref{eqn:Tmat_vac} reduces to 
$\Tmat = 2 \pi a / \redmass (1 + i p_{r} a)$ for a broad Feshbach resonance.

\section{Cross-section}
\begin{figure}[tb]
	\centering
		\includegraphics[width=\columnwidth]{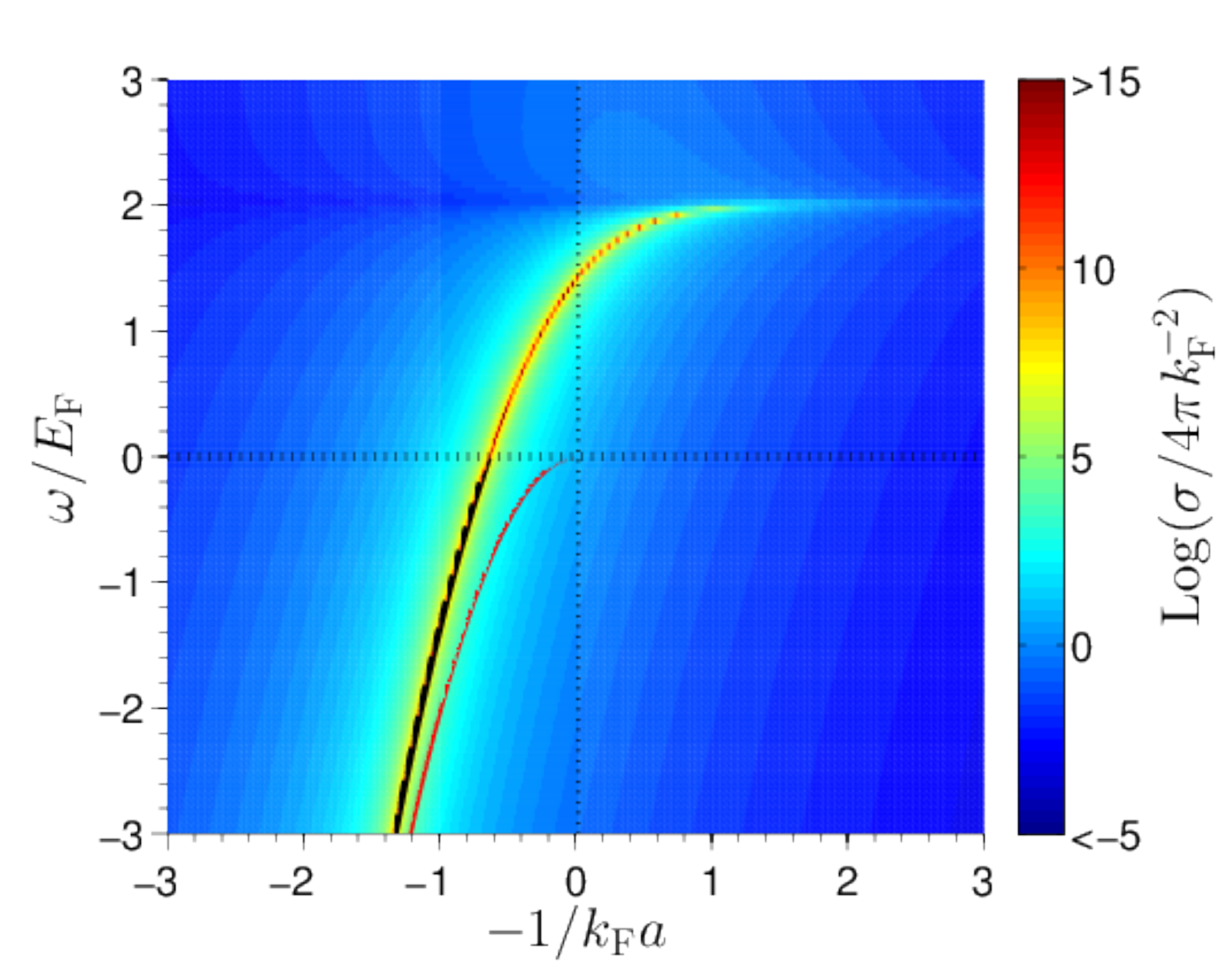}
	\caption{
			The ${\mathbf P}=0$ cross-section $\sigma$ in units of $4 \pi k_\F^{-2}$ as a function of the energy $\omega$ and scattering length $a$
			for  $T = 0.01 \, T_{\F}$, $m_{\downarrow} = m_{\uparrow}$ and $k_{\F} r_{\mathrm{eff}} = 0$. 
			The black line is the numerically obtained  pole of the scattering matrix, while the red line is the molecule pole in a vacuum given 
			by $\epsilon_{\mathrm{B}} = - 1 / 2 \redmass a^{2}$.
			}
	\label{fig:Tmat_kFa_omegaBroadEqual}
\end{figure}

In Fig.~\ref{fig:Tmat_kFa_omegaBroadEqual}, we plot the cross-section $\sigma$ in units of $4 \pi k_{\F}^{-2}$ for a broad resonance with $k_\F r_{\rm eff}=0$ as a function of
scattering energy $\omega$ and scattering length $a$. We have chosen the case of equal masses $m_\uparrow=m_\downarrow$.
The centre of mass momentum is zero, $P=0$, and the temperature is very low with $T=0.01\epsilon_\F $.
 Zero energy corresponds to the bottom of the Fermi sea of the spin $\uparrow$ atoms.
We see a clear pole in the cross section for negative energy on the BEC side of the resonance. It comes from the 
molecule pole which is pushed to higher energies by medium effects, compared to the vacuum molecule energy  
$\epsilon_{\mathrm{B}} = - 1 / 2 \redmass a^{2}$. The ladder approximation employed here provides a qualitatively correct description of the medium energy shift 
which is sufficient for the present purpose, 
although there are quantitative corrections~\cite{Punk2009,Prokofev2008, Massignan_Zaccanti_Bruun}.
The molecule pole is undamped for negative energy whereas it acquires an imaginary part becoming a resonance for $\omega>0$ and $T>0$.
This is because  the molecule can  dissociate into atom pairs with opposite momenta for positive energy.
 For $T=0$, the molecule is   stable up to the energy $2\epsilon_{\F}$  with the pole having no imaginary part, since Fermi blocking 
prohibits dissociation for $\omega\le 2\epsilon_{\F}$. Indeed, a straightforward calculation yields 
\begin{align}
	\Im \left[ \Pi(P=0,\omega)\right] &=
		- \frac{m_r}{2 \pi} \sqrt{2 m_r \omega}
			[ 1 - f_\uparrow ( 2 m_r \omega ) ]
\end{align}
for the imaginary part of the pair propagator describing the decay of the molecule.

  Figure \ref{fig:Tmat_kFa_omegaBroadEqual} clearly demonstrates 
 how the molecule pole on the BEC side smoothly develops into a resonance of the cross-section on the BCS side. 
In the weak coupling BCS limit $k_\F a\rightarrow 0_-$, the resonance position is located at twice the Fermi energy, $\omega=2\epsilon_\F $. 
 This resonance in the cross section  is due to pair correlations, and it is the polaron analogue of the Cooper pole for a balanced system with $n_\uparrow=n_\downarrow$.
  Contrary to the case of a population balanced system where there is a real undamped pole at twice the Fermi energy 
 at the critical temperature for superfluidity, there is no true pole in the BCS-regime of the polaron due to the 
 strong population imbalance. Importantly, we see from Fig.\ \ref{fig:Tmat_kFa_omegaBroadEqual}  
 that the cross section is significantly increased for energy/momenta in the vicinity of the Cooper resonance. This effect is most pronounced close to the 
  unitarity regime whereas it is small in the BEC and BCS limits. 
  The increase in the cross section is important for low temperatures $T\ll \epsilon_\F $, whereas our  numerical calculations  (not shown here) 
  demonstrate  that the Cooper resonance becomes more and more broad with increasing $T$. As expected, it eventually vanishes in the classical limit 
 $T\gg \epsilon_\F $ where the cross section is given by its vacuum value $\sigma_{\rm vac}=m_r^2|{\mathcal T}_{\rm vac}|^2/\pi$.

\begin{figure}[tb]
	\centering
		\includegraphics[width=0.99\columnwidth]{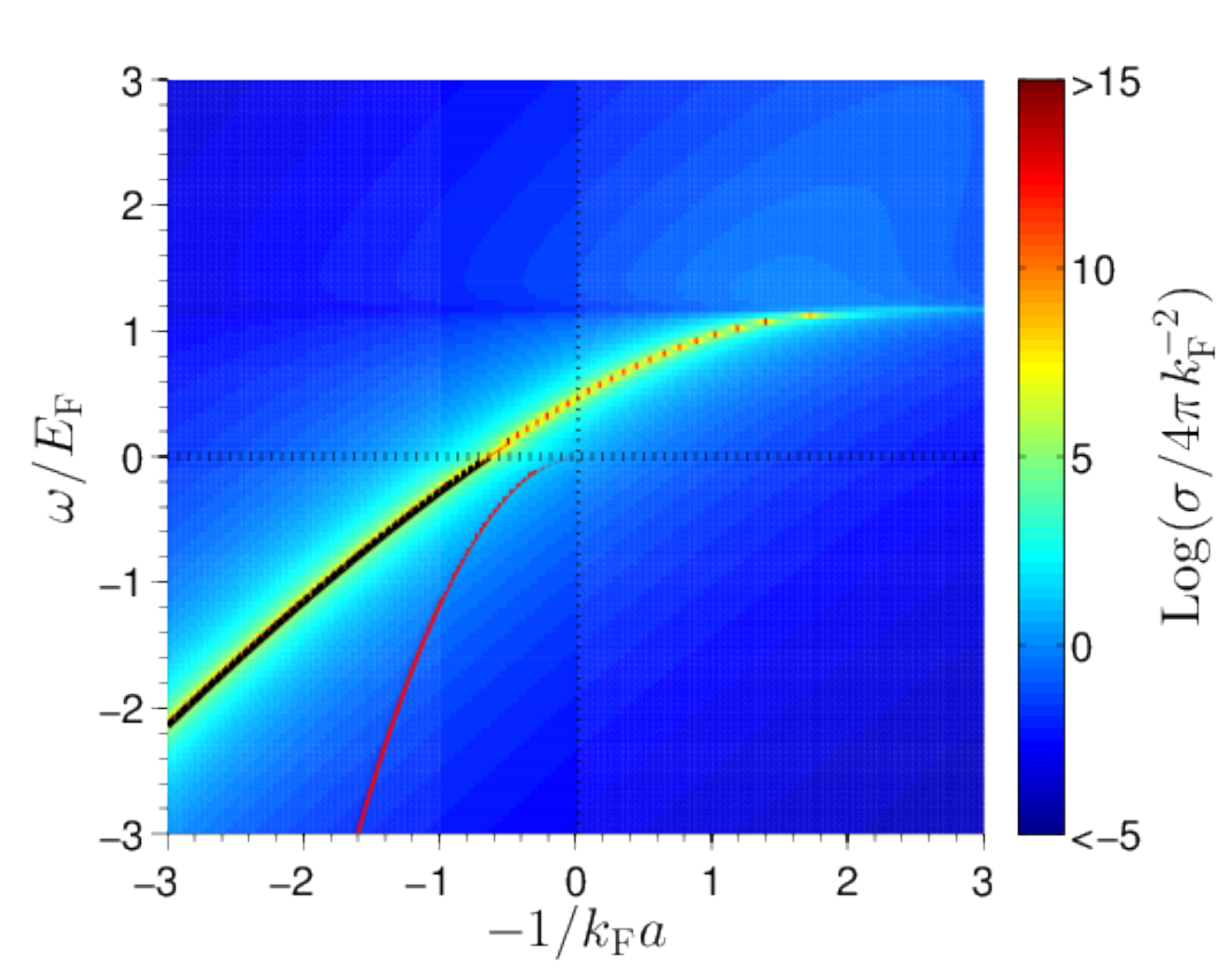}
	\caption{The cross-section $\sigma$ in units of $4 \pi k_\F^{-2}$ as a function of total energy $\omega$ and scattering length $a$ for 
			 $T=0.01 \, \epsilon_\F$, $m_{\downarrow}/m_{\uparrow} = 40/6$ and $k_{\F} r_{\mathrm{eff}} = -1.8$ relevant for the recent experiment using a 
			 $^{40}$K-$^6$Li mixture ~\cite{Kohstall2012}.
			The black line is the numerically obtained  pole of the scattering matrix, while the red line is the molecule pole in a vacuum given 
			by $\epsilon_{\mathrm{B}} = - 1 / 2 \redmass a^{2}$.
			}
	\label{fig:Tmat_kFa_omega}
\end{figure}

The Fermi polaron has recently been investigated experimentally with  $^{40}$K impurity atoms in a Fermi sea of $^6$Li atoms.
The atoms were interacting via a Feshbach resonance  with an effective range 
$k_{\F} r_{\textnormal{eff}}\simeq-1.8$~\cite{Kohstall2012}. 
  In Fig.~\ref{fig:Tmat_kFa_omega}, we therefore  plot the cross section for the parameters of this system: 
 $m_\downarrow/m_\uparrow=40/6$, $k_{\F} r_{\textnormal{eff}}=-1.8$, and a very low temperature of $T=0.01 \, \epsilon_\F $. 
 We see that the qualitative  behaviour is the same as for a broad resonance with mass balance: There is a real molecule pole at negative energy 
on the BEC side which smoothly develops into a Cooper resonance with a non-zero imaginary part 
at positive energy on the BCS side. 
 In the BCS limit $k_{\F} a \rightarrow 0_{-}$, the resonance is located at $\omega = (1 + m_{\downarrow}/m_{\uparrow} ) \epsilon_{\F}$
corresponding to pairing between states at opposite sides of the Fermi surface.
 Comparing Figs.\ \ref{fig:Tmat_kFa_omegaBroadEqual}-\ref{fig:Tmat_kFa_omega} demonstrates   that mass imbalance and a non-zero finite range moves 
the  position of the molecule pole/Cooper resonance to the BCS side as compared to the case of a broad resonance with mass balance. 
Thus, the increase in the cross section due to medium effects is most significant in a region of scattering lengths which is moved towards the BCS side. 
This asymmetry  due to a finite range and mass imbalance effects is consistent 
with what is found for other properties of the polaron~\cite{Massignan_Zaccanti_Bruun}.

\begin{figure}[tb]
	\centering
		\includegraphics[width=0.99\columnwidth]{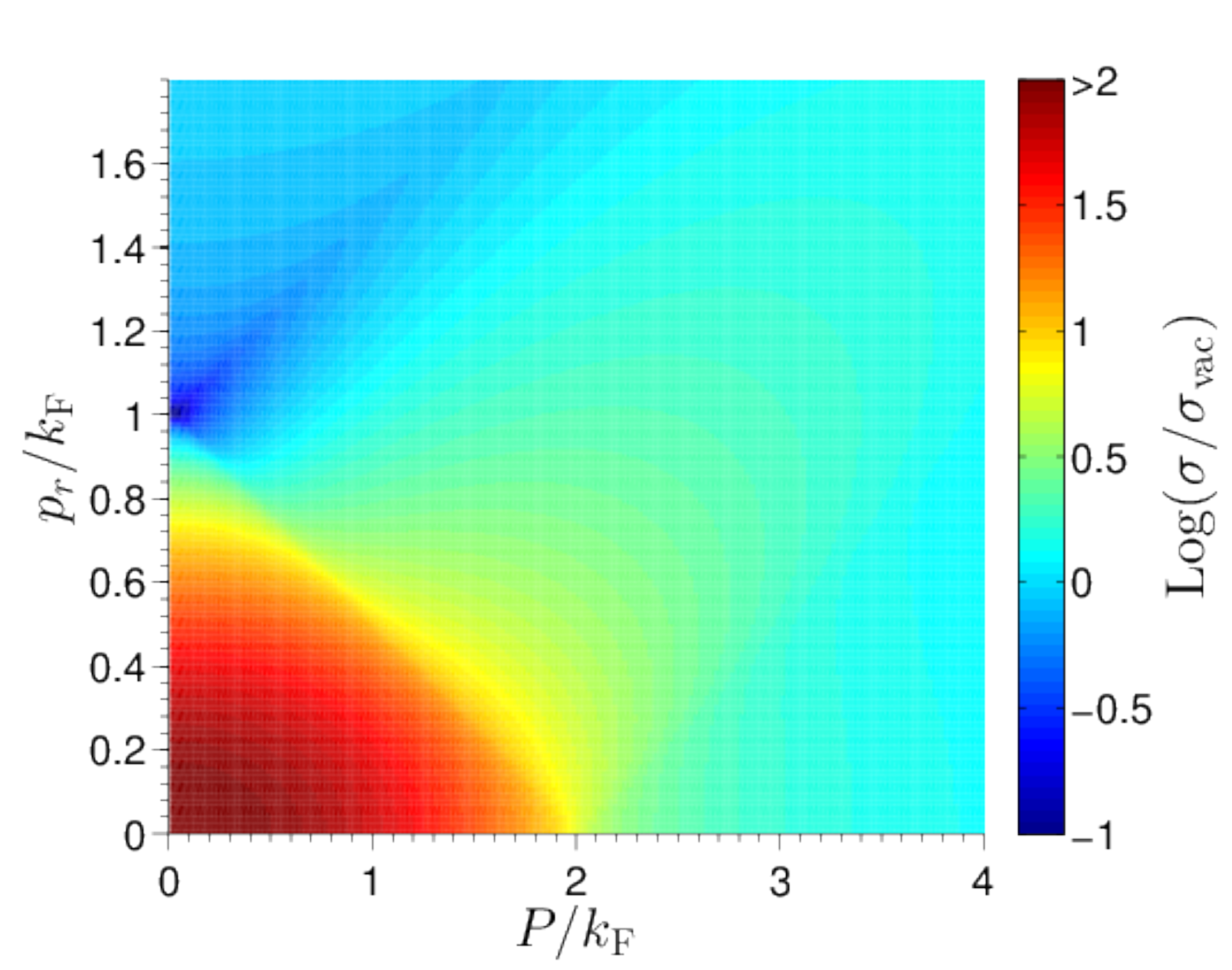}
		\includegraphics[width=0.99\columnwidth]{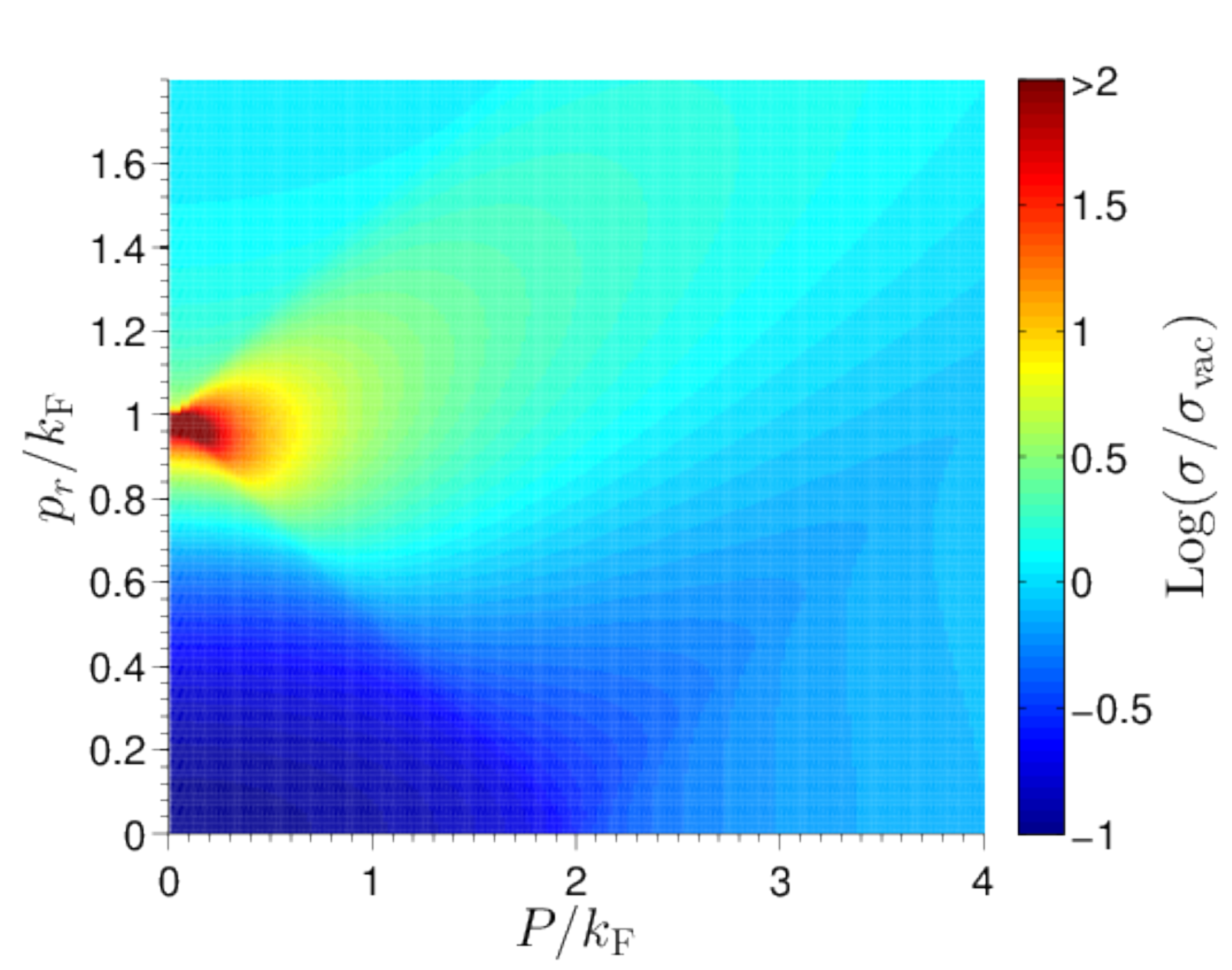}
			\caption{
			The cross-section for $k_\F a = +1$ (top) and $k_\F a = -1$ (bottom) relative to vacuum cross section for 
			$T = 0.01 \; T_{\F}$, $m_{\downarrow} = m_{\uparrow}$ and $k_{\F} r_{\textnormal{eff}} = 0$.
}
	\label{fig:Tmat_P_prBroad_T010}
\end{figure}

We plot in Fig.~\ref{fig:Tmat_P_prBroad_T010} the cross section 
 $\sigma$ in units of the vacuum cross section $\sigma_{\rm vac}$
as a function of the scattering center-of-mass  and  relative momenta for a broad resonance with $m_\uparrow=m_\downarrow$ and 
 $ k_{\F} a = \pm 1$. The temperature is $T=0.01 \, T_{\F}$.
The cross section  is significantly increased compared to its vacuum value for momenta corresponding to energies close  to the 
Cooper resonance. For $k_{\F}a=1$, this gives rise to a large region of enhanced scattering for momenta $\lesssim k_\F $, since  
 the molecule pole/resonance is moved to positive energies by medium effects on the BEC side, as seen in Fig.~\ref{fig:Tmat_kFa_omegaBroadEqual}.
On the BCS side with $k_\F a=-1$, we see that the Cooper resonance gives rise to a sharp increase in the scattering rate for $P\simeq0$ and $p_r \simeq k_\F$.
In the weak coupling BCS limit $k_\F a\rightarrow 0_-$, the Cooper resonance is located at $P=0$ and $p_r = k_\F$, which 
 demonstrates that even in this extreme imbalanced case, the largest pairing correlations are for momenta  at opposite sides of  the Fermi surface of the majority particle.
We furthermore see that the Cooper resonance vanishes with increasing centre of mass momentum. There is therefore 
no trace of a possible Fulde-Ferrel-Larkin-Ovchinikov effect  on the cross section for this imbalanced system. Our numerical calculations show that 
the resonance in the cross section becomes sharper with decreasing temperature, but that  it never becomes a real pole since
 there is no Cooper instability in this strongly imbalanced limit.

\begin{figure}[tb]
	\centering
	\includegraphics[width=0.99\columnwidth]%
		{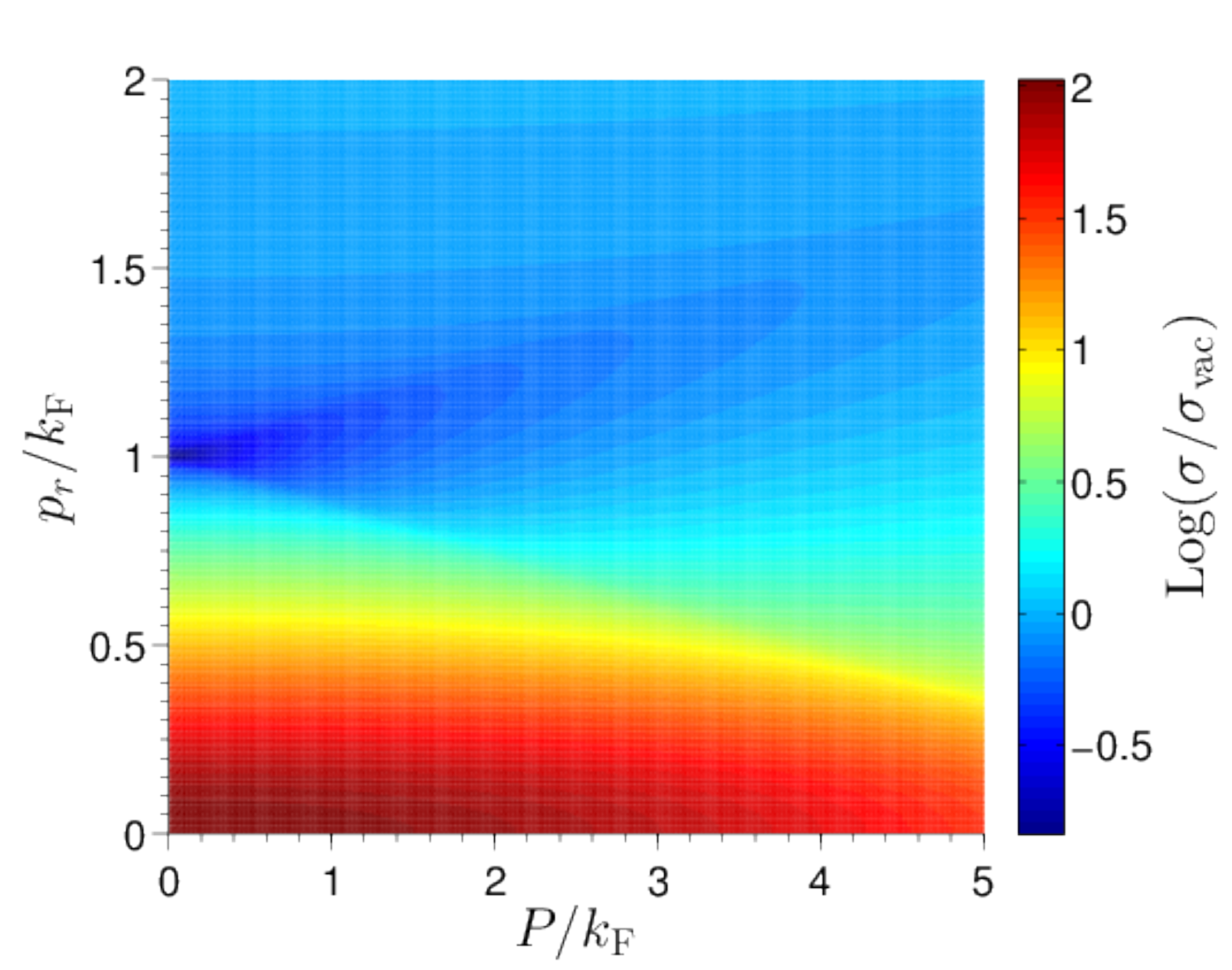}
	\includegraphics[width=0.99\columnwidth]%
		{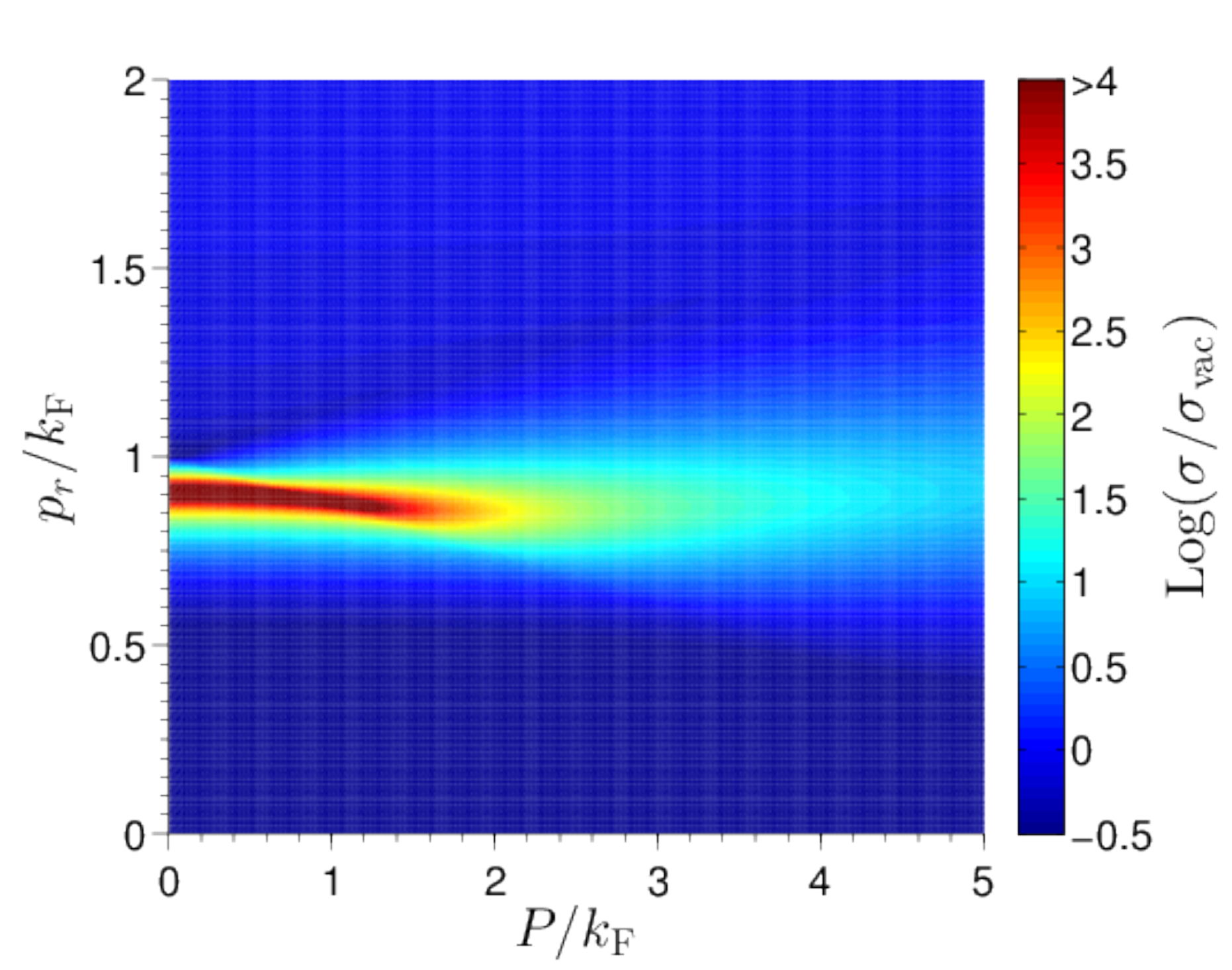}
	\caption{
			The cross-section for $k_\F a = +1$ (top) and $k_\F a = -1$ (bottom) relative to vacuum cross section for 
			$T = 0.01 \; T_{\F}$, $m_{\downarrow} / m_{\uparrow} = 6.6$ and $k_{\F} r_{\textnormal{eff}} = -1.8$.
}
	\label{fig:Tmat_P_prNarrow_T010}
\end{figure}

In Fig.~\ref{fig:Tmat_P_prNarrow_T010}, we plot the cross section as a function of the scattering momenta for the experimentally relevant case 
of a $^{40}$K impurity atom in a Fermi sea of $^6$Li atoms  taking $k_\F a=\pm 1$ and $T=0.01 \, \epsilon_\F $.
The physics is again qualitatively the same as for the case of a broad resonance with mass balance. In particular, we see that the Cooper resonance 
on the BCS side is  located at $P\simeq 0$ and $p_r\simeq k_\F$.
This shows that the strongest pair correlations are  still located on opposite sides of the majority Fermi sea, even for a large mass imbalance
as we already noted above.
The cross section is increased much more on the BCS side for the case of a 
  finite range and mass imbalance compared to  a broad resonance with mass balance, as can be seen by comparing the scales in 
Figs.\  \ref{fig:Tmat_P_prBroad_T010}-\ref{fig:Tmat_P_prNarrow_T010} (bottom). This is again consistent with the general observation
 that a finite range and mass imbalance moves medium effects toward the BCS side.

\section{Collision rate}
 We  now analyse how the increase in the cross section due to pair correlations manifests itself in the collision rate of the polaron. 
  This rate is obtained from  the appropriate momentum average of the cross-section   given in Eq. (\ref{Lifetime2a}), 
  and  we will calculate it as a function of momentum and temperature.

\begin{figure}[tb]
	\centering
		\includegraphics[width=0.99\columnwidth]{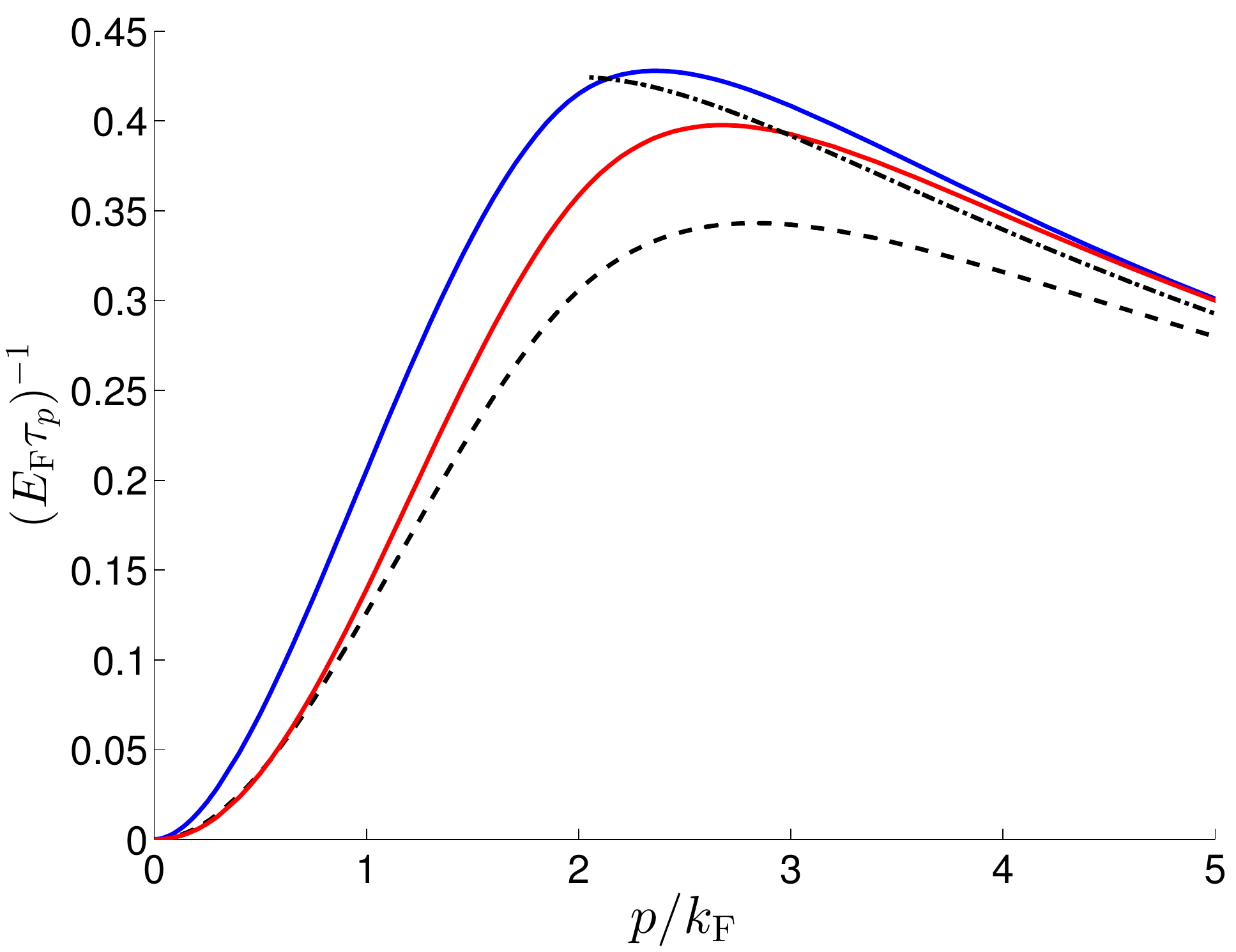}
		\includegraphics[width=0.99\columnwidth]{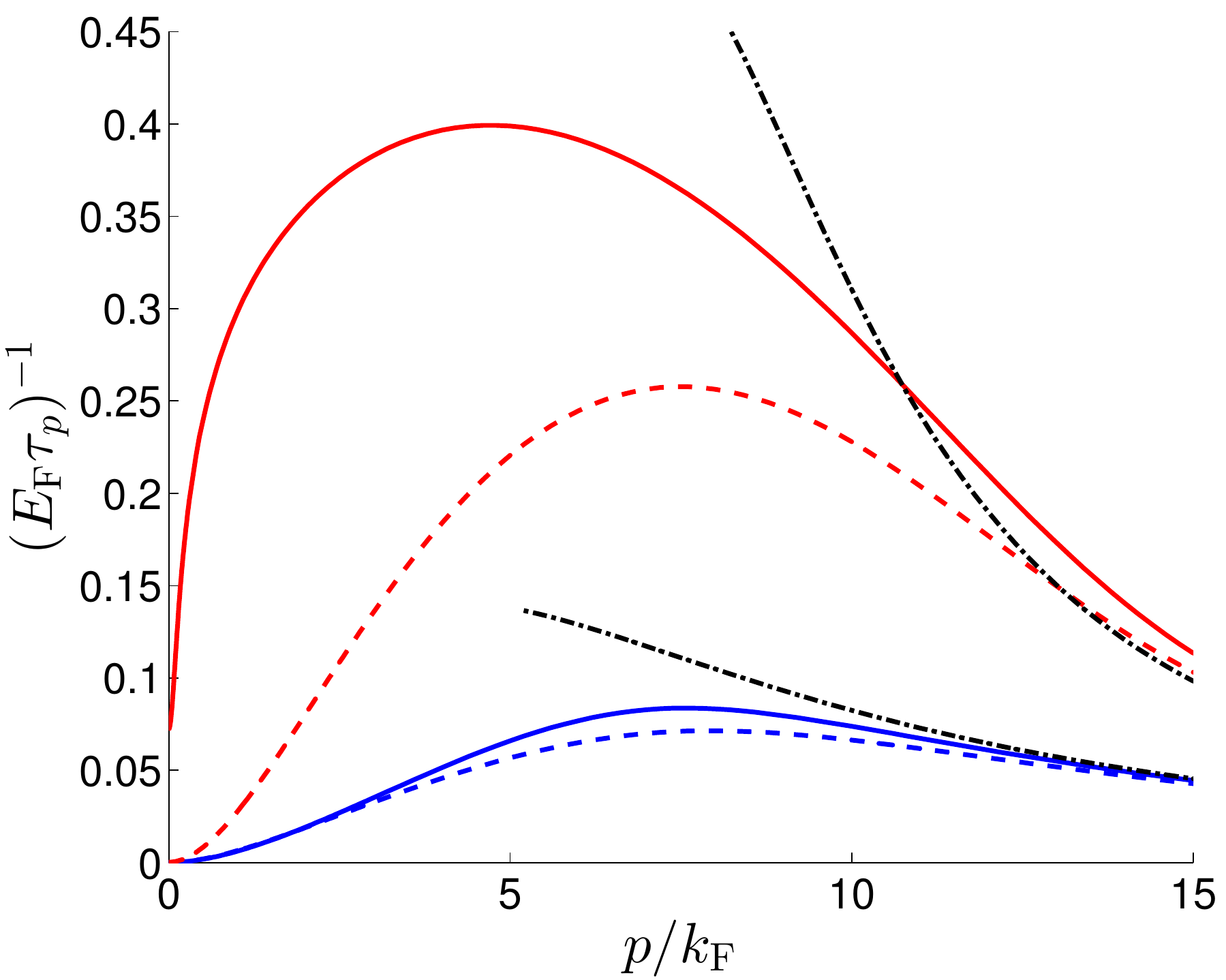}
	\caption{ 
			The collision rate of  the polaron as a function of momentum for $T = 0.01 \; \epsilon_\F$ and for $k_{\F} r_{\textnormal{eff}} = 0$ with 
			equal masses $m_{\downarrow} = m_{\uparrow}$ (top), and the case relevant to the $^6$Li-$^{40}$K mixture with 
			 $k_{\F} r_{\textnormal{eff}} = -1.8$ and $m_{\downarrow} / m_{\uparrow} = 6.6$ (bottom).
			 In both figures, the solid blue line is for $k_{\F} a = +1$, the solid red line is for $k_{\F} a = -1$, 
			 dashed lines are the vacuum rate, and the dash-dotted lines are the $p \gg k_{\F}$ limit given by Eq.\ (\ref{Highp}).
			}
	\label{fig:CollRate_kFa_1pdep}
\end{figure}

 In \cref{fig:CollRate_kFa_1pdep}, we plot the collision rate in units of $\epsilon_\F$ as a function of  momentum for $k_\F a=\pm 1$ for a very low temperature $T=0.01\,\epsilon_\F$.
 We show both the case of a broad resonance with mass balance, and the experimentally relevant case corresponding 
 to a $^{40}$K impurity atom colliding with a Fermi sea of $^6$Li atoms. 
 Consider first the case of a broad resonance with mass balance. 
 Comparing with the "vacuum" collision rate defined as the rate obtained using the vacuum cross section in Eq.\ (\ref{Lifetime2a}), 
  we see that medium effects indeed increase the scattering rate  due to pair correlations. The effect is largest on the BEC side.  
  Note that the vacuum cross section is the same for $k_\F a=\pm 1$ for 
  a broad resonance. Consider next the narrow resonance for which  there is no symmetry between vacuum cross section at 
  $k_\F a=\pm 1$. Figure \ref{fig:CollRate_kFa_1pdep} shows that the significant increase in the collision rate due to pairing correlations is now largest on the BCS side with $k_\F a=-1$. 
  This difference in the medium effects on the collision rate between the broad and the narrow resonance  is consistent with what we found for the cross section. 
We note that pair correlations  have been shown to lead to a similar increase in the 
collision rate for a population balanced system, which strongly affects the damping of  collective modes~\cite{Riedl2008,Baur2013,Urban2014}, 
the shear viscosity and the spin diffusion constant~\cite{BruunSmith2005,Enss2012,Enss2013}.

  We also see that the medium effects are most significant for low momenta, 
  whereas the scattering rate approaches its classical value for high momenta. This is  as expected, since Fermi blocking becomes 
insignificant for large momenta. The collision integral (\ref{Lifetime2a}) can in fact be solved analytically in the limit $p\gg k_\F,\sqrt{2m_\uparrow T}$ yielding
\begin{align}
\frac{1}{\tau_p} &=\frac {4\pi n_\uparrow}{m_\downarrow}
 \frac{a^2p}{\left( \frac{\left( \frac{m_\uparrow}{M} \right)^2
      p^2 r_\mathrm{eff} (a -a_\mathrm{bg}) - 2}{\left( \frac{m_\uparrow}{M} \right)^2
      p^2 r_\mathrm{eff} a_\mathrm{bg} (1 -
      a_\mathrm{bg} / a) - 2} \right)^2 + \left(
      \frac{m_\uparrow}{M} a p \right)^2 }\nonumber\\
      &=\frac {4\pi n_\uparrow}{m_\downarrow}
 \frac{a^2p}{1+\left(
      \frac{m_\uparrow}{M} a p \right)^2},
\label{Highp}
\end{align}
where  the last line holds for a broad resonance. From Fig.\ \ref{fig:CollRate_kFa_1pdep}, we see that the collision rates approach
 this expression for high momenta confirming the accuracy of our numerics.

\begin{figure}[tb]
	\centering
		\includegraphics[width=0.99\columnwidth]{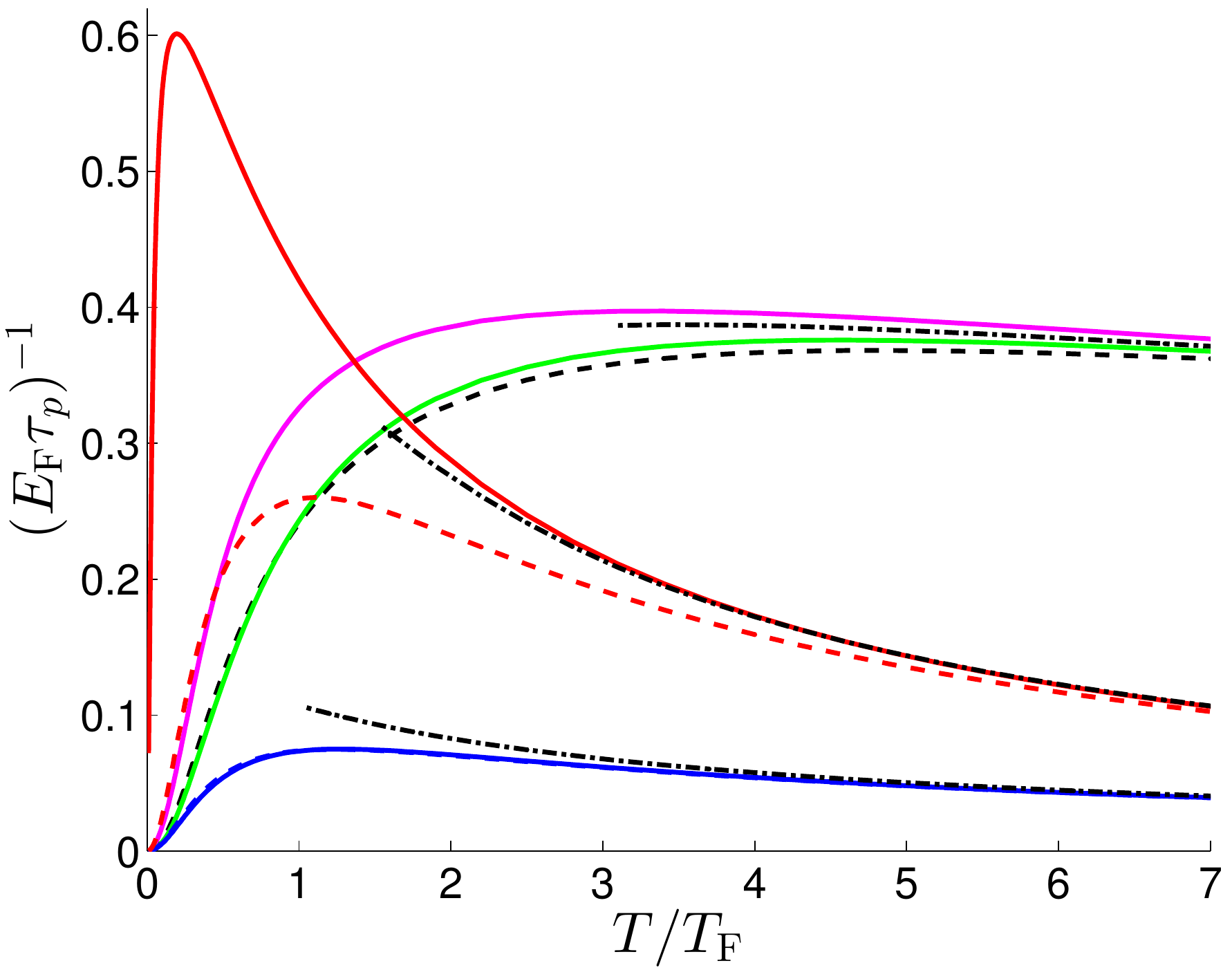}
	\caption{ 
			The collision rate of the polaron as a function of temperature for $p = 0$.
			 The dashed lines are the vacuum rate, the dash-dotted lines are the $p \gg k_{\F}$ limit.
			 The remaining lines are for $k_{\F} r_{\textnormal{eff}} = 0$ with 
			 equal masses $m_{\downarrow} = m_{\uparrow}$ and $k_{\F} a = +1$ (magenta) and $k_{\F} a = -1$ (green),
			 and the case relevant to the $^6$Li-$^{40}$K mixture with $k_{\F} r_{\textnormal{eff}} = -1.8$, 
			 $m_{\downarrow} / m_{\uparrow} = 6.6$ and $k_{\F} a = +1$ (blue) and $k_{\F} a = -1$ (red).
			}
	\label{fig:CollRate_kFa_1Tdep}
\end{figure}

The temperature dependence of the collision rate is plotted in \cref{fig:CollRate_kFa_1Tdep} for  zero momentum in units of $\epsilon_\F $. 
We have as before taken $k_\F a=\pm 1$ and analysed a broad resonance with 
mass balance as well as  the experimentally relevant   case of a  $^{40}$K atom in a  $^6$Li Fermi sea.  
Again, we see by comparing with the value obtained using the vacuum cross section $\sigma_{\rm vac}$ that pairing correlations significantly increase 
the collision rate for low temperatures. Consistent with the results above, medium effects are most significant on the BEC side for the wide resonance whereas they are 
most signifcant on the BCS side for the  narrow resonance. 

For high temperatures $T\gg \epsilon_\F $, the effects of pairing correlations vanish and the scattering rate 
approaches its classical value. The integral in (\ref{Lifetime2a}) can be solved in the classical limit, and we obtain 
 \begin{equation}
	\frac{1}{\tau_{p }} =
	2 n_{\uparrow} \sqrt{ \frac{2 T}{\pi m_{\uparrow} } } \bar\sigma
\end{equation}
for  $p\ll\sqrt{m_\downarrow^2T/m_\uparrow}$. Here 
\begin{gather}
	\bar\sigma=8\pi a^2\int _0^\infty dx\frac{x^3e^{-x^2}}{\left(\frac{Tr_{\rm eff}(a-a_{\rm bg})x^2-2T_aa^2}{Tr_{\rm eff}a_{\rm bg}(1-a_{\rm bg}/a)x^2-2T_aa^2}\right)^2+\frac{T}{T_a}x^2}\nonumber\\
	=
	8\pi a^2\int _0^\infty dx\frac{x^3e^{-x^2}}{1+\frac{T}{T_a}x^2}
	\label{sigmabar}
	\end{gather}
is an effective cross section, 
and $T_a$ is defined as
\begin{equation}
T_a=M^2/2m_\uparrow m_\downarrow^2a^2.
\end{equation}
The last line in Eq.\ (\ref{sigmabar}) holds for  a broad resonance. For a broad resonance, Eq.\ (\ref{sigmabar}) yields $\bar\sigma=4\pi a^2$ for the  
effective cross section  in the weak coupling limit $T_a\gg T$, and $\bar\sigma=2\pi M^2/m_\uparrow m_\downarrow^2T$ in the strong coupling limit $T_a\ll T$. This gives 
\begin{equation}
	\frac{1}{\tau_{p \downarrow}} =
			4 n_{\uparrow} \sqrt{ \frac{2 \pi }{ m_{\uparrow} } } \times
				\left\{
					\begin{gathered}
						2 a^{2} \sqrt{T} \quad \text{for} \quad T_a\gg T
					\\
						\frac{m_{\uparrow}}{\redmass^{2}} \frac{1}{\sqrt{T}} \quad \text{for} \quad T_a\ll T.
					\end{gathered}
				\right.
	\label{eqn:Lifetime_High_T}
\end{equation}
Figure  \ref{fig:CollRate_kFa_1Tdep} shows as expected that  the collision rates approach the classical value for $T\gg \epsilon_\F $ 
given by Eq.\ (\ref{eqn:Lifetime_High_T}), confirming the accuracy of our numerics.

\section{Experimental observation}
We now briefly discuss how  the collision rate of the impurity atom can be detected  experimentally. 
One method is to use RF spectroscopy to 
Rabi flip  the impurity atom between an internal state  interacting with the majority Fermi sea and a non-interacting internal  state. 
The collisions of the impurity atom when it is in the interacting state cause decoherence  which results in a  damping of the Rabi 
oscillations. This damping has in fact already 
been observed experimentally in the $^{40}$K-$^6$Li mixture~\cite{Kohstall2012}. Using this, one can extract the collision rate from the measured decoherence rate.  
The collision rate can also be measured using Bose-Fermi mixtures. One could create a    Bose-Einstein condensate of impurity 
atoms  in a finite momentum state which  then collides with the majority Fermi sea.  The advantage of this method is that the polaron is then in a definite 
momentum state. However, there will be non-trivial effects on the scattering rate due to the presence of the BEC which must be examined. 

\section{Conclusions}
We have analysed the scattering rate of an impurity atom with non-zero momentum in a Fermi sea of majority atoms. 
The cross section for the Feshbach resonance mediated interaction was calculated using a microscopic multichannel theory, 
which includes finite range and medium effects. We demonstrated that  correlations between states with opposite momenta
significantly increase the cross section. These correlations are due to molecule formation at negative energy on the BEC side of the resonance,
 which smoothly connects to a 
 Cooper pair resonance on the BCS side for energies and momenta comparable to the 
Fermi energy. They are  the polaron analogue of  superfluid pairing for the corresponding population balanced system. We demonstrated that the 
pair correlations lead to a considerable increase in the low temperature scattering rate, for impurity momenta smaller than 
or comparable to the Fermi momentum of the majority atoms. The effects of  mass imbalance and a finite range of the interaction  were shown to be significant for  
the experimentally relevant $^6$Li-$^{40}$K mixture. Finally, we discussed how the scattering rate of an impurity atom can be measured using RF spectroscopy or 
Bose-Fermi mixtures. 

\acknowledgements
We acknowledge useful discussions with Marko Cetina.
We would like to acknowledge the support of the Hartmann Foundation via grant A21352 and the Villum Foundation via grant VKR023163.

\bibliography{bibliography}

\end{document}